\documentclass[%
preprint,
 amsmath,amssymb,
 aps,
]{revtex4-1}

\usepackage{graphicx}
\usepackage{dcolumn}
\usepackage{bm}

\begin{document}


\title{Investigating the electronic structure of MSi (M = Cr, Mn, Fe \& Co) and calculating \textit{U\textsubscript{eff}} \& \textit{J} by using cDFT
}

\author{Paromita Dutta}
 \altaffiliation{dutta.paromita1@gmail.com}
 \affiliation{%
School of Basic Sciences, Indian Institute of Technology Mandi, Kamand, Himachal Pradesh-175005, India}%
\author{Sudhir K. Pandey} 
\affiliation{%
School of Engineering, Indian Institute of Technology Mandi, Kamand, Himachal Pradesh-175005, India
}%

\date{\today}

\begin{abstract}

We have investigated electronic energy band structures and partial density of states of intermetallic compounds \textit{viz.} CrSi, MnSi, FeSi and CoSi, by using density functional theory (DFT). CrSi \& MnSi, FeSi and CoSi have metallic, indirect band gap semiconducting (band gap $ \sim$ 90 meV) and semi-metallic ground state, respectively. On studying the band structures while going across the series Cr-Co, the occupied bands around the Fermi level are getting narrower while the unoccupied bands are getting wider. Similarly, band edge in partial density of states is shifting away from the Fermi level due to increased hybridizations. The effective mass of holes for FeSi is found to be much larger than that of electrons, giving rise to positive Seebeck coefficient and negative Hall coefficient, which is consistent with experimental results. For different ionic states of 3\textit{d}-metal, the values of \textit{U\textsubscript{eff}} and \textit{J} are evaluated by using constrained DFT method. \textit{U\textsubscript{eff}} (\textit{J}) for $2^{+} $ ionic state across the series are $\sim $ 3.3 eV ($\sim $0.65 eV), $\sim $ 3.7 eV ($\sim $0.72 eV), $\sim $ 4.4 eV ($\sim $ 0.82 eV) and $\sim $ 4.5 eV ($\sim $ 0.87 eV). $\lambda$ and \textit{J} are also calculated by considering Yukawa form of Coulomb interaction. $\lambda $ values for $2^{+} $ ionic state along the series are $\sim $ 1.97 a.$u^{-1}$, $\sim $ 2.07 a.$u^{-1}$, $\sim $ 2.07 a.$u^{-1}$ and $\sim $ 2.34 a.$u^{-1}$. 4\textit{s} electrons are found to be contributing more in screening the 3\textit{d} electrons as compared to 4\textit{p} electrons of 3\textit{d} metals.

\end{abstract}

\maketitle


\section{Introdution}

Silicon is one of the crucial material in electronic industry. Transition metals form a very interesting group of materials with silicon due to their peculiar magnetic and electrical properties. Beacuse of this, transition metal monosilicides MSi (M= Cr, Mn, Fe and Co) with B20 cubic structure of spacegroup P2\textsubscript{1}3 have been studied for their various properties during last decades. For example, CrSi is a pauli paramagnetic metal [1,2], MnSi is an itinenarant helimagnetic metal for T$<$30 K [3], FeSi is a paramagnetic kondo insulator [4] and CoSi is a diamagnetic metal [1,2,5-6], with resistivities of $ \sim$ 2.4 $\Omega$-cm, $ \sim$ 2.0 $\Omega$-cm, $ \sim$ 3.4 $\Omega$-cm and $ \sim$ 1.5 $\Omega$-cm, respectively, at room temperature [1]. All these compounds are exhibiting wide range of properties although their lattice parameters change a very little. This reflects the urge of understanding the electronic structural properties exhibited by silicon and the transition metal. And their study is quite challenging due to the presence of correlation effects of \textit{d} electrons and the screening effects. Although a very limited study of transition metal monosilicides have been done individually. But till now, we are unaware of any systematic study for the electronic states for these intermetallic compounds experimentally and theoretically. Moreover, experimentally it is found that FeSi shows large thermoelectric power i.e., a positive Seebeck coefficient $ \alpha$ at the same time a negative Hall coefficient [7]. These different signs of ($ \alpha$) and Hall coefficient indicating the dominant charge carriers to be \textit{p}-type and \textit{n}-type. These interesting properties of FeSi are not well understood. So, it will be quite interesting to study this aspect.   
 
  DFT has successfully given eminent results for many materials [8-15]. In light of this, many properties of these intermetallic compounds have been studied by using DFT method [16-19]. However, DFT studies the correlation effect exhibited by \textit{d} electrons present in compounds in a very limited way. Thus, to have a firm understanding of correlations in a more effective way, DFT+\textit{U} and beyond studies are made. For carrying out these studies, two important parameters are required, and they are on-site Coulomb interaction \textit{U}, and Hund's coupling \textit{J}, where \textit{U} and \textit{J} both correspond to correlations existing within orbitals. In these studies, one appropriate value of \textit{U} and \textit{J} are choosen in order to match with experimental results for giving a qualitative reasoning for their physical properties. For instance, studies for various physical properties of FeSi and MnSi have been done using DFT+\textit{U} method , where \textit{U} is taken as parameter [20,21]. For example, from DFT+\textit{U} study for FeSi done by Anisimov \textit{et al.} [20] has shown that the study of anomalous properties of FeSi depends on the choice of \textit{U}, such as \textit{U} was choosen $\geq$ 3.0 eV for first order transition from singlet semiconductor to ferromagnetic metal. Similarly, Collyer \textit{et al.} have varied \textit{U} value over range 0$\leq$\textit{U}$\leq$9.5 eV, and has found that for \textit{U}$\leq$6.8 eV, the magnetic moment is going with the experimental results [21]. However, DFT+\textit{U} works on CoSi and CrSi are not known as per our knowledge. Thus, it will be appreciable if one can calculate a material specific values of \textit{U} and \textit{J} of these intermetallic systems for studying their physical properties in an extensive manner. Also, in solids screening effect is much larger as when compared to atomic case, due to which screening plays an important role in deciding the values of \textit{U\textsubscript{eff}} and \textit{J} [22,23] for compounds. So, it will be fascinating to see the role of screening while studying the behavior of \textit{U} and \textit{J} for these intermetallic compounds as well.
  
   In this work, we report a detailed investigation of electronic structure for 3\textit{d}-metal monosilicides \textit{viz.} CrSi, MnSi, FeSi and CoSi, by use of DFT under local density approximation (LDA) [24]. By studying their band structures and partial density of states(PDOS) plots, a systematic study of their electronic structures is done. It is found that CrSi \& MnSi, FeSi and CoSi are metallic, semiconducting with an indirect gap $\sim$ 90 meV and semi-metallic ground state. Also the effective mass of holes are found to be greater than that of electrons, which gives rise to the positive Seebeck cofficient and negative Hall coefficient as observerd experimentally. With PDOSs, an insight of occurring hybridisations are realized. For different ionic states, we have calculated \textit{U\textsubscript{eff}}, \textit{J} and screening length $ \lambda$ values separately for all of these intermetallic compounds which are given in Table II. and IV. Although, we have shown two approaches for computing \textit{J} and it is found that both \textit{J}'s values are approximately same. Also, the trends exhibited by these parameters have been discussed.
   
\begin{table}
\caption{\footnotesize{Lattice parameters (a) and atomic coordinates (x). Both M (M = Cr, Mn, Fe and Co) and Si atoms are located at the 4(a)-type sites in the simple-cubic unit cell, with position coordinates at (x,x,x).}}
\label{tab.1}
\begin{center}
\setlength{\tabcolsep}{40pt}
\footnotesize
\begin{tabular}{lccr}
\hline
Compound  & a & x\textsubscript{M} & x\textsubscript{Si} \\
    & (\AA) &  & \\
\hline 
\hline   
CrSi & 4.629 & 0.136 & 0.846 \\
MnSi & 4.558 & 0.137  & 0.845 \\
FeSi &  4.493 & 0.136   &  0.844  \\
CoSi & 4.438 & 0.140 & 0.843 \\
\hline
\end{tabular}
\end{center}
\end{table}

\begin{table}
\caption{\footnotesize{Evaluated \textit{U\textsubscript{eff}} and \textit{J} from cDFT for MSi (M = Cr, Mn, Fe and Co) compounds for their different ionic states. The bracket values correspond to \textit{J}.}}
\label{tab.II}
\begin{center}
\setlength{\tabcolsep}{25pt}
\footnotesize
\begin{tabular}{lccccr}
\hline
Compound  & 1\textsuperscript{+} & 2\textsuperscript{+} & 3\textsuperscript{+} & 4\textsuperscript{+} \\
   & \textit{U\textsubscript{eff}}(\textit{J}) & \textit{U\textsubscript{eff}}(\textit{J}) & \textit{U\textsubscript{eff}}(\textit{J})& \textit{U\textsubscript{eff}}(\textit{J})  \\
  & (eV) & (eV) & (eV) & (eV)\\  
\hline 
\hline   
CrSi & 1.9(0.56) & 3.3(0.65) & 5.3(0.85) & 7.5(1.01)\\
MnSi & 3.7(0.72) & 3.7(0.72)  & 5.8(0.84) & 7.7(1.05)\\
FeSi & 4.4 (0.82)& 4.4(0.82)   &  6.2(0.72) & 7.9(0.97)\\
CoSi & 4.5 (0.87)& 4.5(0.87) & 6.5(0.97) & 7.9(0.98)\\
\hline
\end{tabular}
\end{center}
\end{table}

\begin{table}
\caption{\footnotesize{Number of electrons in the occupied region of 3\textit{d}-metal monoscilides evaluated from partial DOS within LDA. 4\textit{p} and 3\textit{d} are the higher energy states of 3\textit{d}-metal atom and Si atom, respectively.}}
\label{tab.3}
\begin{center}
\setlength{\tabcolsep}{20pt}
\footnotesize
\begin{tabular}{lccccccr}
\hline
Compound &  & 3\textit{d}-metal &   & & Si \\
  & 4\textit{s} & 3\textit{d} & 4\textit{p} &  3\textit{s} & 3\textit{p} & 3\textit{d}\\
\hline 
\hline   
CrSi &  0.48 & 4.4 & 0.36 & 0.6 & 0.7 & 0.70\\
MnSi &  0.50 & 5.6 & 0.40 & 0.6 & 0.69 & 0.69\\
FeSi &  0.4 & 6.6 & 0.44 & 0.6 & 0.7 & 0.68\\
CoSi &  0.5 & 7.6 & 0.48 &  0.6 & 0.7 & 0.68\\
\hline
\end{tabular}
\end{center}
\end{table}

\begin{table}
\caption{\footnotesize{Evaluated $\lambda$'s and \textit{J}'s values corresponding to each evaluated \textit{U\textsubscript{eff}} values for MSi (M = Cr, Mn, Fe and Co) compounds at their different ionic states. The bracket values correspond to \textit{J}.}}
\label{tab.4}
\begin{center}
\footnotesize
\setlength{\tabcolsep}{20pt}
\begin{tabular}{lccccr}
\hline
Compound  & 1\textsuperscript{+} & 2\textsuperscript{+} & 3\textsuperscript{+} & 4\textsuperscript{+} \\
    & $\lambda$ (\textit{J}) & $\lambda$ (\textit{J})   & $\lambda$ (\textit{J})  & $\lambda$ (\textit{J})  \\
     & a.u\textsuperscript{-1} (eV)&  a.u\textsuperscript{-1} (eV) & a.u\textsuperscript{-1} (eV) & a.u\textsuperscript{-1} (eV)\\
\hline 
\hline   
CrSi & 2.84 (0.54) & 1.97 (0.72) & 1.34 (0.87) & 0.94 (0.98)\\
MnSi & 2.07 (0.78) & 2.07 (0.78)  & 1.41 (0.95) & 1.06 (1.03)\\
FeSi & 2.07 (0.89) & 2.07 (0.89)  &  1.55(1.02) & 1.22 (1.10)\\
CoSi & 2.34  (0.94)& 2.34  (0.94)& 1.72 (1.10) & 1.43 (1.17)\\
\hline
\end{tabular}
\end{center}
\end{table}

\section{Computational Details}

\par{The calculations for electronic structures of MSi (M = Cr, Mn, Fe and Co) have been performed here. These calculations can be divided into two segments. First segment involves DFT calculation and other segment as for the computation of \textit{U\textsubscript{eff}} for each compound separately. All these calculations are performed by using full-potential linearized augmented plane-wave (FP-LAPW) method as accomplished by WIEN2k code [25]. LDA is taken as exchange correlation functional [24]. The experimentally observed structural parameters and atomic positions are taken from literature [26-28] and they are given in Table I. The muffin-tin sphere radii of 2.45 and 1.84 bohr for M (M = Cr, Mn, Fe and Co) and Si sites, respectively with 10$\times$10$\times$10 k-point mesh size have been used for all calculations. For first segment, spin-unpolarized calculations are performed.}

\par{The spin-polarized calculation is performed within constrained DFT (cDFT) for the computation of \textit{U\textsubscript{eff}} and \textit{J} for M 3\textit{d} atom in MSi. The method for computing its numerical value is proposed by Anisimov \textit{et.al} [23]. In this method, within a constructed supercell, a hoping term (3\textit{d} orbital of one atom is connected with rest of the obitals of remaining atoms) is set to zero. By altering the numbers of electrons in non-hybridising 3\textit{d} shell, \textit{U\textsubscript{eff}} for correlated 3\textit{d} shell can be evaluated by the following formula }

 \begin{eqnarray} 
\textit{U\textsubscript{eff}} = \epsilon_{3\textit{d}\uparrow}\Big(\frac{n+1}{2},\frac{n}{2}\Big)-\epsilon_{3\textit{d}\uparrow}\Big(\frac{n+1}{2},\frac{n}{2}-1\Big) 
 - \epsilon_{\textit{F}}\Big(\frac{n+1}{2},\frac{n}{2}\Big)+\epsilon_{\textit{F}}\Big(\frac{n+1}{2},\frac{n}{2}-1\Big)
\end{eqnarray}

\par{where $\epsilon_{3\textit{d}\uparrow}$ and $\epsilon_{\textit{F}}$ are the spin up 3\textit{d} eigenvalue and the Fermi energy. The implementation of the given procedure is done by Madsen \textit{et.al} [29], achieved through above mentioned code [25]. Then, the procedure for evaluating \textit{U\textsubscript{eff}} for these 3d metal monosilicides is followed as given in Lal \textit{et al.} paper [30]. cDFT has given quite an appreciable results for studying correlations exhibited by \textit{d} electron systems [31,32]. Similar procedure is followed for calculating \textit{J} values by using the following formula [33]:

\begin{eqnarray} 
-\textit{J} = \epsilon_{\textit{d}\uparrow}\Big(\frac{n+1}{2},\frac{n-1}{2}\Big)-\epsilon_{\textit{d}\downarrow}\Big(\frac{n+1}{2},\frac{n-1}{2}\Big) 
 \end{eqnarray}

 The evaluated \textit{U\textsubscript{eff}} and \textit{J} for all these compounds are given in tabular form in Table II. at different ionic states of individual 3\textit{d} metal.}

\section{Results and Duscussion}

\begin{figure}[!b]
  \begin{center}
    \includegraphics[width=5in]{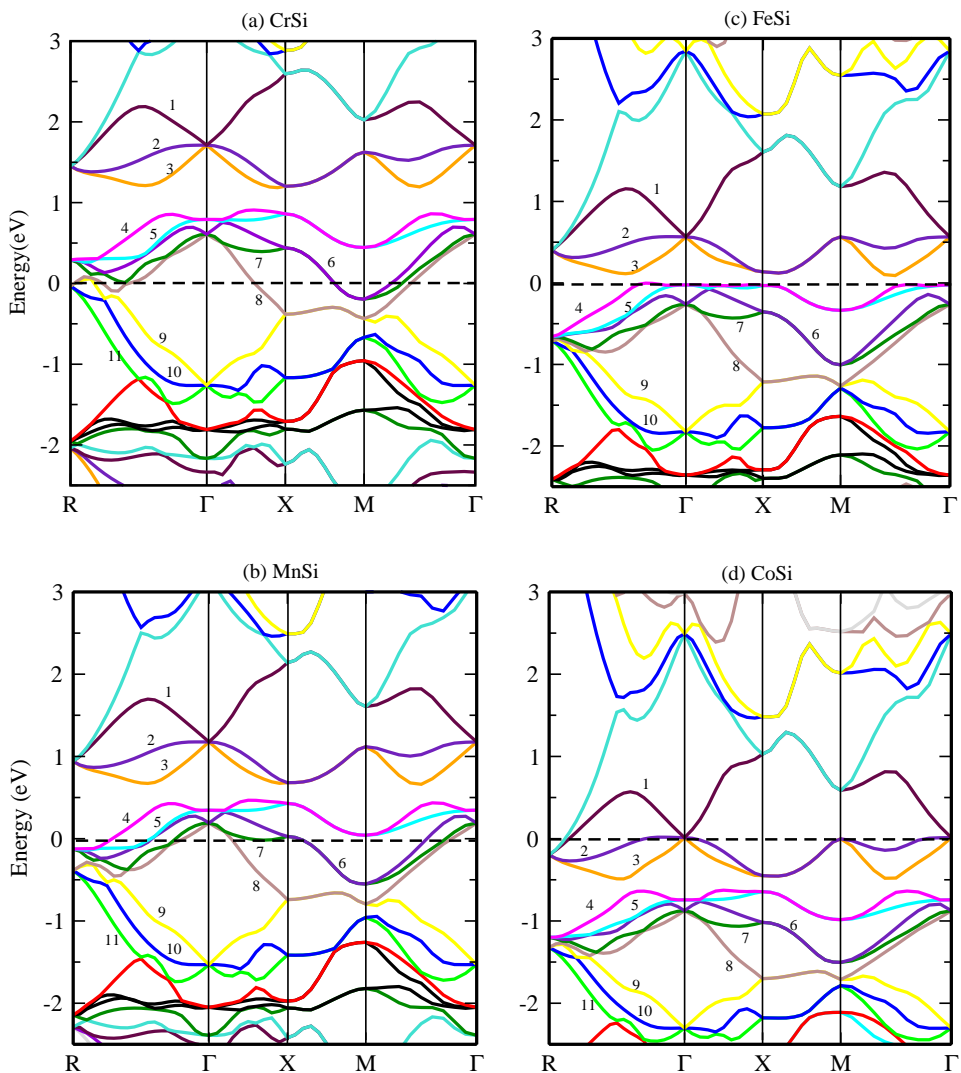}
  \end{center}

  \caption{\small{(Colour online) Band structures of 3\textit{d}-metal monosilicides i.e., (a) CrSi, (b) MnSi, (c) FeSi and  (d) CoSi for the energy range $ \sim$ -2.5 eV to $ \sim$ 3.0 eV are shown. Zero energy corresponds to the Fermi level. }}
 
\end{figure}

\begin{figure}[!b]
  \begin{center}
    \includegraphics[width=5in]{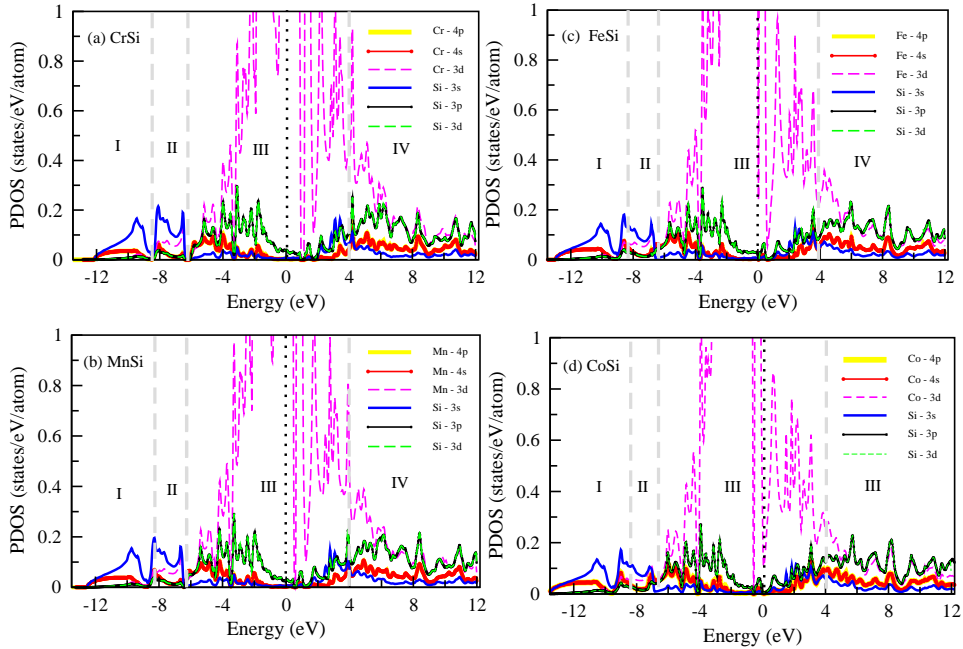}
  \end{center}

  \caption{\small{(Colour online) Partial density of states (PDOS) of M 4\textit{s}, M 3\textit{d}, M 4\textit{p}, Si 3\textit{s}, Si 3\textit{p} and Si 3\textit{d} states of MSi (M = Cr, Mn, Fe and Co) within DFT. Zero energy corresponds to the Fermi level. }}
 
\end{figure}

 The results are discussed in three sections. In the first and the second part, a detailed investigation of the electronic structure exhibited by these intermetallic compounds are given for the whole series in a comparative way by studying both the band structures and partial density of states (PDOS). Third part involves the trend shown by the calculated values of \textit{U\textsubscript{eff}}, \textit{J} and $ \lambda$ for these intermetallic compounds when discussed along the series with the role of screening in determinig the values of these parameters.
\\
\subsection{\textbf{ Band structure}}

 In Fig. 1(a)-1(d), band structures of MSi(M = Cr, Mn, Fe and Co) for the energy range $ \sim$ -2.5 eV to $ \sim$ 3.0 eV are plotted. In these plots, 11 bands are identified for the investigation to be done for these intermetallic compounds. In Fig. 1(a), it is  observed that band 8 is crossing the Fermil level while band 7 is just touching the Fermi level along R-$\Gamma$ direction. Similary, along the direction $\Gamma$-X, band 8 is the only one crossing the Fermi level while along X-M direction, band 6 is crossing. And in the end, along M-$\Gamma$ direction, bands 6-8 are all crossing the Fermi level. Then, just above the Fermi level, two more bands i.e., 4 and 5 are lying where they are crossing each other. So, this crossing of the Fermi level by bands, is directly showing the existence of metallic nature of CrSi which is in accordance with the experimentally observed metallic state [1]. Moreover, on moving to the higher scale of energy, it is seen that now bands 3 and 4 are much farther than each other with a gap of $\sim $ 0.27 eV. Also, along R-$\Gamma$ direction, bands 1-3 are well separated while from $\Gamma$-M, bands 2 and 3 are crossing each other, although band 1 is well separated from the other two. The bandwidth of band 1 is of $ \sim$ 1.19 eV. similarly for band 10, which is lying on the much lower side of energy scale with a bandwidth of $\sim$ 1.29 eV. 
 
 Here it will be interseting to see the effect if we replace the Cr metal with Mn, Fe or Co metal. Owing to this, two effects will be observed: (i) the increase in effective nuclear charge, and (ii) filling of unoccupied bands.
The first effect can be understood as when Cr metal is changed with any one of them as mentioned above, the number of electrons and protons will increase. As a result the effective nuclear charge increases i.e., each orbital of the atom will feel the pull more effectively and they will come close to the nucleus. But depending on the spatial spread of each orbital they will feel the pull differently. Although, due to the increase in the effective nuclear charge, a consequent decrease in their ionic radii is expected, which will further result in  the decrease of the lattice parameter i.e., atoms will be much closer now as it can be seen from the Table I. where experimentally observed lattice parameters are tabulated. Now it is expected to have the probability of overlappings to be more as their interatomc distances are decreasing which simply indicates that the bandwidths of bands are expected to increase always. But at the same time a competitive nature of interactions will come into picture i.e., as the spatial spreading of the orbitals have already contracted but when the amount of contraction of orbitals is more than the half of the interatomic distances, then after overlapping the bandwidth is expected to decrease and when the amount of contraction is less than the half of the interatomic distances, the respective bandwidths are expected to increase. Thus, based on these situations, the respective bandwidths of the bands may increase and decrease accordingly. The second effect i.e., the presence of extra electrons, will just fill bands from the unoccupied region. Due to which, the Fermi level will shift towards the higher side of energy scale. 

 Now, we will start with CrSi, where Cr metal will be replaced by Mn metal. And it will be quite exciting to observe the respective changes due to the effects as discussed above can actually be observed by MnSi or not. The impacts are observed by comparing Fig. 1(a) and 1(b). On observing Fig. 1(b), it is noted that bands i.e., 4-8 are now crossing the Fermi level along R-$\Gamma$ direction unlike CrSi case and for the directions $\Gamma$-X, X-M and M-$\Gamma$, there's no such change in the bands except that along $\Gamma$-X direction band 7 is now touching the Fermi level. When we move from Cr metal to Mn metal, there will be increment of 4 electrons per unit cell. Each band accommodates two electrons per k-points and one unit cell will contribute 4 extra electrons. With the fact that number of k-points are equal to the number of unit cells, two more bands are expected to be  filled by these 4 extra electrons. 
Because of the electron filling in the bands, it is expected to have shift in the Fermi level towards the higher side of energy scale, the direct consequence of second effect as discussed above. This is what observed from the figure where bands 4 and 5 seem to have crossed the Fermi level. In this way, it is again indicating the metallic nature of MnSi as it is found experimentally [1].
The number of bands crossing the Fermi level has increased which is indicating the increased DOS around the Fermi level. Because of this, the resistivity is expected to decrease which is exactly to what experimentally found [1]. 
The rest of the bands are behaving in the same manner as found in CrSi with similar shapes of the bands due to the fact that they exhibit same crystal structure type. The first effect as mentioned above now can be realised here when we compute the individual bandwidths of band 3 and 4. It is found that the gap between bands 3 and 4 has now reduced to $\sim$ 0.20 eV, this is due to the increase in the bandwidths of bands 3 and 4.

 When we replace the Mn metal with Fe metal, the consequences of the effects can be discussed by comparing Fig. 1(c) with 1(b). In considering this replacement of metal, there will be increment of 4 more electrons in a unit cell. As it is evident from Fig. 1(b) that bands 7 and 8 are almost filled. So, bands 4 and 5 are now expected to be filled with extra electrons with a shift in the Fermi level. This is exactly we can notice from Fig. 1(c) where bands 4-8 are exactly inside the Fermi level. Band 3 is above the Fermi level with a bandwidth of $\sim$ 0.47 eV while the bandwidth of band 4 has increased to $\sim$ 0.65 eV. Consequently, there is a formation of a small indirect band gap of $\sim$ 90 meV. This outcome is fairly in aggreement with experimental and theoretical results where it is found to be $\sim$ 60 meV  and $\sim$ 110 meV [7,34]. Thus, it is indicating that FeSi is an indirect bandgap semiconductor. For this material, a very interesting behavior is observed i.e., on measuring Hall coefficient for FeSi a negative value is found which is indicating the dominating charge carriers to be of \textit{n}-type. However, on measuring the Seebeck coefficient for this material, a large positive value is coming, indicating the thermoelectric property is dominated by \textit{p}-type charge carriers [7]. By studying the shown band structure of FeSi in Fig. 1(c), one can understand these interesting outcomes. In intrinsic semiconductor, the number of electrons and holes are normally said to be equal. And FeSi being intrinsic semiconductor, it is expected to have zero Hall coefficient due to the equal number of positive and negative charge carriers. However, this equivalence of charge carriers is expected only at 0 K due to the fact that the chemical potential lies at the center of the band gap as evident from Eq. 3 [35].
 
 \begin{equation}
\mu = \varepsilon\textsubscript{v} + \frac{1}{2} E\textsubscript{g} + \frac{3}{4} k\textsubscript{B} T \ln(m_{v}/m_{c})
\end{equation}

where, where $\varepsilon\textsubscript{v}$ is the energy of the electron at the top of the valence band (VB), $ E\textsubscript{g} $ is the energy of band gap, $m_{v}$ and $m_{c}$ are the effective mass of the charge carrier in valence and conduction band (CB), respectively. Since measuremnets are carried out at finite temperatures, so it is important to see the behavior of $\mu$ at finite temparatures. According to the expression given in Eq. 3, $ \mu$ has dependency in both temparature and  effective mass of charge carriers. Thus, at finite temperature, there is an expectation for the chemical potential to shift. Now when we look at the band structure of FeSi in Fig. 1(c), the top of the valence band is flat while bottom of the conduction band has curvature.
This will lead to the large value of effective mass of holes (m\textsubscript{h}\textsuperscript{*}) than effective mass of electrons (m\textsubscript{e}\textsuperscript{*}) [36,37]. In consequence, chemical potential is expected to shift upwards towards the conduction band. This clearly shows the case of electron doping i.e., dominant charge carriers to be of n-type at finite temperature. This is exactly in accordance with the experimental Hall coeefecient result for FeSi [7]. Futhermore, Seebeck coefficient has dependence on  both the effective mass of charge carriers as well as the electron number density according to the expression given in Eq. 4, which as follows [38]:

\begin{equation}
\alpha = (8 \Pi^{2} k_{B}^{2} / 3 e h^{2}) m^{*}T(\Pi/3n)^{2/3}
\end{equation}

where, $k_{B}$ is the Boltzmann constant and $e$ is the electronic charge, $m^{*}$ and $n$ are the effective mass of charge carrier and carrier density, respectively. In FeSi, the number of electrons has increased due to which the value of $n$ will be more than the value of $p$. Although the value of $n$ is greater here, still the value of m\textsubscript{h}\textsuperscript{*} will be much larger when compared to $n$. As a result, primarily  m\textsubscript{h}\textsuperscript{*} will contribute to the large and positive value of $\alpha $. Thus, indicating the thermoelectric property dominated by  p-type charge carrier in FeSi as observed experimentally. Thus, in this way, band structure of FeSi is able to provide a qualitative explanation of the above mentioned two different aspects as shown by FeSi. 

 On replacing the Fe metal with Co metal, again there will be addition of 4 more electrons. Accordingly, bands will be filled and as expected bands 2 and 3 should go inside the Fermi level and this can be predicted on looking at Fig. 1(c). It is again evident from Fig. 1(d) where bands 2 and 3, both have completely crossed the Fermi level while band 1 has crossed over the Fermi level along R-$\Gamma$ direction while rest part of the band 1 is just touching the Fermi level. This is clearly indicating the almost semi-metallic nature of CoSi [1]. After studying all these plots from Fig. 1(a)-1(b), it is quite interesting to note that all those bands which are crossing the Fermi level on filling with electrons, their respective bandwidths are decreasing while those bands which are above the Fermi level are showing the increasing bandwidths. For example, band 1 has shown the trend of increasing bandwidths (i.e., $ \sim$1.19, $ \sim$1.33 , $ \sim$1.41 and $ \sim$1.49)  while band 10 has shown the decreasing trend (i.e., $ \sim$1.29, $ \sim$1.20, $ \sim$1.19 and $ \sim$1.02). Although band 3 has first shown the increasing trend (i.e., $ \sim$0.33 and $ \sim$ 0.51) before crossing the Fermi level and decreasing trend  (i.e., $ \sim$0.47 and $ \sim$ 0.46). this behavior will be quite helpful in understanding the transport behavior of these compounds.

\subsection{\textbf{ Partial density of states}}

In Fig. 2, PDOSs of M 4\textit{s}, M 3\textit{d}, M 4\textit{p}, Si 3\textit{s}, Si 3\textit{p} and Si 3\textit{d} states are shown. For identifying the contributions from various orbitals, the study has made by considering small energy windows. Accordingly, 4 energy windows are categorized here as: (I) $\sim $ -13.5 eV to $\sim $ -8.3 eV, (II)
$\sim $ -8.3 eV to $\sim $ -6.1 eV (III) $\sim $-4.0 eV to $\sim $4.0 eV, and (IV)$\sim $4.0 eV to $\sim $12.0 eV. From Fig. 2(a), one can observe that in the (I) energy window, Cr 4\textit{s}, Cr 3\textit{d}, Si 3\textit{s} and Si 3\textit{d} states are participating with contributions as $\sim $ 22.6\%, $\sim $ 7.0\%, $\sim $ 65.0\% and $\sim $4.0\%, respectively. Similarly, in the (II) energy window, Cr 4\textit{s}, Cr 3\textit{d}, Si 3\textit{s} and Si 3\textit{d} states are participating with contributions as $\sim $ 10.0\%, $\sim $ 30.0\%, $\sim $ 53.3\% and $\sim $8.7\%, respectively. Meanwhile, in (III) energy window, almost all the contribution is coming from Cr 3\textit{d} with contribution as $ \sim$ 87.0\%. 
 However, in the (IV) energy window, the contribution from states Cr 3\textit{d}, Si 3\textit{s} and Si 3\textit{d} are almost equal. Energy window (III), can be realised by looking at the band structure plot for CrSi in Fig. 1(a). On comparing the Fig. 1(a) and 2(a), it is found that the energy range in which zero DOS is seen, is the same region which lying in between the gap exhibited by the bands 3 and 4. Moreover, all the identified bands in Fig. 1(a) are mostly contributed by Cr 3\textit{d} orbital, as around the Fermi that has the maximum contribution as seen in Fig. 2(a).

 Now as we move from CrSi to CoSi along the series, there are two changes seen in the partial DOSs from Fig. 2(a)-2(d), and they are as follows: (i) shifting of all PDOS peaks towards the lower side on the energy scale, and (ii) the intensity of peaks are changing. For example consider the peak corresponding to Si 3\textit{s}, which is at first peaked $ \sim$ -9.5 eV for the case of CrSi, then it is moved to $ \sim$ -9.8 eV when it is for MnSi, then to $ \sim$ -10.0 eV, and finally to $ \sim$ -10.47 eV. Similar behaviour can be seen for all other orbitals. This can be understood as when the interatomic distances are small, the hybridization 
increases resulting in the increased separation of band edge to move away from the Fermi level. Moreover, number of electrons are increasing which gives rise to filling of bands with these increased electrons. As a result, PDOSs peaks of \textit{d} states shift on the lower side of the energy scale along the series. This results in shifting of Fermi level on the higher side on the energy scale. Now consider the intensities exhibited by hybridizing orbitals in region I and II, here again a substantial change is found in them from region to region. This can be understood as the atoms come closer, the occurence of hybridisations will be more. Because of which the sharing of electrons between the participating orbitals in hybridizations will change. Inconsequence of this, there will be redistribution of density of states between the orbitals. Thus, the probablity of finding electron on the main orbital to which other orbitals are hybridizing will decrease. For example, when we move along the series, only contribution from Si 3\textit{s} state has decreased by $ \sim$ 11.3\% while contributions from rest of the states have increased for region I. This simply indicating that  Si 3\textit{s} orbital is the main one to which other orbitals are hybridizing. Due to which the probability of finding electron has decreased in Si 3\textit{s} orbital, and accordingly its intensity has decreased. 
Similar, behaviour can be seen for region II while moving along the series. Although for this region, now contributions from Si 3\textit{s} and \textit{d} orbitals of transition metal has reduced by $ \sim$ 23.6\% and $ \sim$18.7\%, respectively. So one can say that for this region, other orbitals are hybridizing with both Si 3\textit{s} states and \textit{d} orbitals of transition metal. Along the series, for region III, intensity of PDOS peaks corresponding to transition metal \textit{d} states are increasing due to the filling of bands with increased electrons number.

\subsection{\textbf{ Investigation of \textit{U\textsubscript{eff}}, \textit{J} and $\boldsymbol{\lambda$ }}}

 The calculations for finding the value of effective Coulomb interaction \textit{U\textsubscript{eff}} and Hund's coupling \textit{J} are dicussed in the follwing section. The suitable value of \textit{U\textsubscript{eff}} and \textit{J} for  the localised 3\textit{d} electrons are computed by using cDFT method. In this method, \textit{d}-linearization energy (E\textit{\textsubscript{d}}) plays a vital role in it. Thus, to find the suitable values of \textit{U\textsubscript{eff}} and \textit{J} for different ionic states of 3\textit{d}-metal, E\textit{\textsubscript{d}} is kept $\sim$ 40 eV above Fermi level. And the corresponding calculated values of \textit{U\textsubscript{eff}} and \textit{J} for MSi (M = Cr, Mn, Fe and Co) are tabulated in Table II. The further validation is done from the PDOSs of the compounds for different ionic states of 3\textit{d}-metal. It is noted that, the contribution of \textit{d} states of PDOSs around the Fermi level for impurity atom is negligible. This suggests that the value of E\textit{\textsubscript{d} }is appropriate for finding the value of \textit{U\textsubscript{eff}} and \textit{J} [30]. Since, we could not find any experimental ionic states for these 3\textit{d} metal of these intermetallic compounds. Therefore, we have calculated \textit{U\textsubscript{eff}} and \textit{J} values for different ionic states  i.e., $1^{+} $, $2^{+} $, $3^{+} $ and $4^{+} $ of transition metals. Generally, ionic states of $2^{+} $, $3^{+} $ and $4^{+} $ for transition metals are expected in different compounds. For these compounds, the bonding is pure ionic in nature, then one expects metal ions to be at $2^{+} $ state. At the same moment, if it is covalent in nature, then depending upon the degree of covalency, one can expect metal ions to have less than $2^{+}$ ionic state. Moreover, for other compounds where transition metals have oxidation states of $3^{+} $ and $4^{+} $, then  corresponding to these states the calculated values of \textit{U\textsubscript{eff}} and \textit{J}, can be taken from this report and these values might be a good starting point for performing other theoretical calculations. Accordingly, we have studied the changes  occurring in the values of these parameters with different ionic states. From Fig. 2, the number of electrons corresponding to M 4\textit{s}, M 3\textit{d}, M 4\textit{p}, Si 3\textit{s}, Si 3\textit{p} and Si 3\textit{d} states are evaluated, and they are tabulated in Table III. 
 
   On looking at Table II., one can find a sequence in which the respective values of \textit{U\textsubscript{eff}} and \textit{J} are increasing along the series i.e., from Cr to Co .
For example, for $1^{+}$ ionic state of metal ion, the value of \textit{U\textsubscript{eff}} (\textit{J}) has increased from $ \sim$ 1.95 eV ($ \sim$ 0.56 ) to $ \sim$ 3.7 eV ($ \sim$ 0.72 ) upto $ \sim$ 4.5 eV ($ \sim$ 0.87 ) across the series.
 To understand the above metioned trend, let us take an example of CrSi. In atomic case of Cr and Si, Cr 3\textit{d}, Cr 4\textit{s}, Si 3\textit{s} and Si 3\textit{p} states have 5, 1, 2 and 2, number of electrons, respectively  whereas in solid case, the number of electrons as calulated for the states Cr 4\textit{s}, Cr 3\textit{d}, Cr 4\textit{p}, Si 3\textit{s}, Si 3\textit{p} and Si 3\textit{d} are found as $\sim $ 0.4, $\sim $ 4.4, $\sim $0.36, $\sim $ 0.6, $\sim $0.7 and $\sim $ 0.7, respectively. This shows that on forming a solid, number of electrons have redistributed among themselves due to hybridizations occurring between orbitals. Moreover, for transtion metals, n\textit{s} and n\textit{p} orbitals extend more than (n-1)\textit{d} orbitals (where n is the principal quantum number) which makes them to be away from the nucleus. Although, most  of the screening faced by 3\textit{d} orbitals are coming from 4\textit{s} and 4\textit{p} orbitals primarily for transition metals [23]. 
This says that 3\textit{d} electrons of Cr metal will be screened by electrons lying in 4\textit{s} and 4\textit{p} orbitals. 
Due to the presence of ligand atom in CrSi, reflecting it to consider the screening from electrons lying in other states of ligand atom. At the same moment, it is known that the extention of 3\textit{s} and 3\textit{p} orbitals are very less, due to which electrons lying in these orbitals will be more confined to the Si atom. Consequently, electrons residing in these orbitals of Si are not expected to contribute much to the screening faced by 3\textit{d} oribtals' electrons of Cr metal ion. Since, number of electrons residing in 4\textit{s} and 4\textit{p} orbitals are almost equal in number as observed from Table III. With the fact that 4\textit{s} orbital is more closer to the 3 \textit{d} orbital, due to which 4\textit{s} will more overlap with 3\textit{d} orbitals. Hence, we can say that the electrons residing in 4\textit{s} orbital will screen 3\textit{d} elelctrons more than electrons residing in 4\textit{p} orbital. Now, on observing the number of electrons in 4\textit{s}  and 4\textit{p} states of transition metal across the series in Table III., it is noted that these numbers of electrons are approximately constant. So, we expect that the screening provided by transition metal 4\textit{s} and 4\textit{p} states may not be changing and thus its effect is not significant while looking for transition metal 3\textit{d} electrons' interactions across the series. 
 
 Furthermore, along the series, number of 3\textit{d} electrons of metal ion are increasing. In consequence of this, the effective distance between the 3\textit{d} electrons will decrease. This will make 3\textit{d} electrons to interact more strongly and consequently the values of \textit{U\textsubscript{eff}} and \textit{J} are expected to increase across the series, and which is evident from Table II. It is important to note that the values of \textit{U\textsubscript{eff}} and \textit{J} for $ 1^{+}$ and $2^{+} $ ionic states of MnSi to CoSi are found to be equal due to the unchanged number of electrons in their 3\textit{d} orbitals. Similarly, there is another trend where \textit{U\textsubscript{eff}} and \textit{J} values are increasing from 1\textsuperscript{+} to 4\textsuperscript{+} oxidation states of the same 3\textit{d}-metal. Like for CrSi, one can find that when we move from Cr 1\textsuperscript{+} to Cr 2\textsuperscript{+} till Cr 4\textsuperscript{+} , the value of \textit{U\textsubscript{eff}} (\textit{J}) have increased from $ \sim$ 1.95 eV ($ \sim$ 0.56) to $\sim $ 3.3 eV ($ \sim$ 0.65) upto $\sim $ 7.5 eV ($ \sim$ 1.01), respectively. This can be realized as on increasing the ionic states, the number of valence \textit{d} electrons keep on decreasing. Due to which the effective nuclear charge will increase, making the \textit{d} electrons to feel the pull more strongly. This may contract the 3\textit{d} orbitals more in comparision to 4\textit{s} and 4\textit{p} orbitals resulting in the decrease of the overlappings occurring between 4\textit{s} and 4\textit{p} orbitals with 3\textit{d} orbitals. As a result,\textit{d} electronic interactions will increase, and consequently, the values of \textit{U\textsubscript{eff}} and \textit{J} will increase with the increase in ionicity of 3\textit{d}-metal. The arguments gievn above are all made on qualitative basis.
  
 There is another way of calculating the value of \textit{J} by considering Yukawa potential instead of Coulomb potential. Using the value of \textit{U\textsubscript{eff}}, Yukawa screening $\lambda$ is evaluated. Then this $\lambda$ value is used for calculating the higher orders of Slater intergrals $F^{2}$ and $F^{4}$ which are further related with \textit{J} by the following expression given in Eq. 4:
 
 \begin{equation}
 \textit{J} = \frac{F^{2} + F^{4}}{14}
 \end{equation}
 
For computing $\lambda$ and \textit{J}, we have used eDMFT code as implemented by Haule \textit{et al.} [39]. The evaluated values of $\lambda$ and \textit{J} are tabulated in Table IV. Although, two approaches have been used for calculating \textit{J}, but their values are found to be approximately same as evident from Table II. and IV. Also the trends shown by both \textit{J}s are similar. Likewise, $\lambda$ too has shown decreasing trend while moving along the series with increasing ionicity of 3\textit{d}-metal as evident from Table IV. Across the series, number of \textit{d} electrons are increasing, thereby an increase in electronic interactions will be there, resulting in decrease in $\lambda$ values across the series. Similarly, along the  different ionic states of 3\textit{d}-metal, there will be decrase in the number of valence electrons, resulting in less overlapping between 4\textit{s} and 4\textit{p} with 3\textit{d} orbitals. In consequence, screening will decrease substancially, thereby a decreasing trend is seen for $\lambda$ values.

\section{Conclusion}
 
 Here, we present a consistent account of electronic structures exhibited by 3\textit{d} transition metal monosilicides \textit{viz.} CrSi, MnSi, FeSi and CoSi by using DFT. Detailed investigation is done in a comparative way by studying band structure and PDOSs as obtained. By studying band structures, CrSi \& MnSi, FeSi and CoSi are found to be metallic, semiconducting with an indirect band gap of $ \sim$ 90 meV and semi-metallic, respectively, which is consistent with experimental results. Across the series CrSi to CoSi, bandwidths of occupied bands closer to the Fermi level are decreasing while unoccupied bands' bandwidths are increasing. Simiarly, band edge of PDOSs are shifting away from the Fermi level, indicating more overlappings between orbitals.
From band structure of FeSi, it is found that the effective mass of holes are much larger than that of electrons, indicating a positive Seebeck coefficient and a negative Hall coefficient, as observed experimentally. For different ionic states of 3\textit{d}-metal, \textit{U\textsubscript{eff}} and \textit{J} are calculated by cDFT method. Similarly, \textit{J} and $ \lambda$ values are calculated for each \textit{U\textsubscript{eff}} are evaluated by considering Yukawa form of Coulomb interaction. It is also found that \textit{U} and \textit{J} values are increasing when one goes along the series as well as with the increasing ionic states of 3\textit{d}-metal while $ \lambda$ value is decreasing in the same order.  
 
\section{Data availability} 

The processed data required to reproduce these findings cannot be shared at this time as the data also forms part of an ongoing study.


\begin{thebibliography}{0}

\bibitem{Shinoda} D. Shinoda and  S. Asanabe, J. Phys. Soc. Japan {\bf 21}, 555 (1966).

\bibitem{Wernick} J. H. Wernick, G. K. Wertheim and G. Sherwood, Mat. Res. Bull.  {\bf 7}, 1431 (1972).

\bibitem{Grigoriev} S. V. Grigoriev, S. V.  Maleyev, A. I. Okorokov, Y. O. Chetverikov, P. Boni, R. Georgii, D. Lamago, H.  Eckerlebe andK. Pranzas, Phys. Rev. B  {\bf 74}, 214414 (2006).
  
\bibitem{Paschen} S. Paschen, E. Felder, M. A. Chernikov, L. Degiorgi, H. Schwer, H. R. Ott, D.P. Young, J. L. Sarrao and Z. Fisk,   Phys. Rev. B {\bf 56}, 12916 (1997).

\bibitem{Han} J. -G Han and F. Hagelberg, Chem. Phys.  {\bf 263}, 255 (2001).

\bibitem{Nakanishi} O. Nakanishi, A. Yanase and A. Hasegawa, J. Magn. Magn. Mater. {\bf 15}, 879 (1980).

\bibitem{Imada} M. Imada, A. Fujimori and Y. Tokura, Rev. Mod. Phys. {\bf 70}, 1039 (1998).  

\bibitem{Hohenberg} P. Hohenberg and W. Kohn, Phys. Rev. B {\bf 136}, 864 (1964).

\bibitem{Kohn} W. Kohn and L. J. Sham, Phys. Rev. A {\bf 140}, 1133 (1965).
 
\bibitem{Gross} R. M. Dreizler and E. K.U. Gross, \textit{Density-Functional Theory: An Approach to the Quantum Many-Body Problem},  (Springer, Berlin, 1990)
 
 \bibitem{Pandey} S. K. Pandey, Phys. Rev. B {\bf 81}, 2010 (035114). 
  

\bibitem{Sonu} S. Sharma and S. K. Pandey, J. Magn. Magn. Mater. {\bf 403}, 1 (2016). 
 
\bibitem{Maiti} S. K. Pandey and K. Maiti, EPL {\bf 88}, 27002 (2009). 

\bibitem{Georg} K. H. Georg, G. K. H. Madsen, K. Schwarz, P. Blaha and D. J. Singh, Phys Rev B {\bf 68}, 125212 (2003). 

\bibitem{Martin} R. M. Martin, \textit{Electronic Structure Basic Theory and Practical Methods} (Cam-bridge University Press, 2004)

\bibitem{Mattheiss} L. F. Mattheiss and D. R. Hamann, Phys. Rev. B {\bf 47}, 13114 (1993).

\bibitem{Imai} Y. Imai, M. Mukaida, K. Kobayashi and T. Tsunoda, Intermetallics {\bf 9}, 261 (2001).  

\bibitem{Guevara} J. Guevara, V. Vildosola, J. Milano and A. M. Llois, Phys. Rev. B {\bf 69}, 184422 (2004).  
  
\bibitem{Altintas} B. Altintas, J. Phys. Chem. Solids {\bf 72}, 1325 (2011).  

\bibitem{Ezhov} V. I. Anisimov, S. Yu Ezhov and I. V. Solovyev, Phys. Rev. Lett. {\bf 76}, 1735 (1996). 

\bibitem{Collyer} Collyer R. D. and  Browne D. A., Physica B {\bf 403}, 1420 (2008). 
  
\bibitem{Herring} C. Herring, \textit{Magnetism}, edited by G. T. Rado and H. Suhl (Academic, New York, 1966), Vol. IV

\bibitem{Anisimov} V. I. Anisimov and O. Gunnarsson, Phys. Rev. B {\bf 43}, 7570 (1991). 

\bibitem{Jones} R. O. Jones and O. Gunnarsson, Rev. Mod. Phys. {\bf 61}, 689 (1989). 

\bibitem{Blaha} P. Blaha, K. Schwarz, G. K. H. Madsen, D. Kvasnicka and J. Luitz, \textit{WIEN2k, An Augmented Plane Wave Plus Local Orbitals Program for Calculating Crystal Properties } (Vienna University of Technology Vienna, 2001)

\bibitem{Kemi} B. Bor\'{e}n B. and Ark. Kemi, Min. Geol. {\bf 11A}, 1 (1934). 

 
\bibitem{Jeong} Jeong T. and Pickett W. E., Phys. Rev. B {\bf 70}, 075114 (2004). 

\bibitem{Boren} B. Bor\'{e}n and Ark. Kemi, Min. Geol. {\bf 11A}, 1 (1933). 
 

\bibitem{Novk} G. K. H. Madsen and P. Novk , Europhys. Lett. {\bf 69}, 777 (2005). 
 
\bibitem{Lal} S. Lal S. and S. K. Pandey, Phys. Lett. A {\bf 381}, 2117 (2017) 
  
\bibitem{S} S. Lal and S. K. Pandey, EPL {\bf 117}, 37002 (2017).


\bibitem{Dutta} P. Dutta, S. Lal and S. K. Pandey, {arXiv:1711.03812v1}.
  
\bibitem{DMFT}
{\footnotesize{http://hauleweb.rutgers.edu/tutorials/Constrained+DMFT.html}}
   
\bibitem{Fu} C. Fu, M. P. C. M. Krijn and S. Doniach, Phys. Rev. B {\bf 49}, 2219 (1994)
       
   
\bibitem{b.u} Ashcroft N. W. \and Mermin N. D., \textit{Solid State Physics}, edited by Crane D. G. (Saunders College Publishing, New York, 1976), Vol. 239

\bibitem{Maurya} Singh S., Maurya R. K. \and Pandey S. K., J. Phys. D: Appl. Phys. {\bf 49}, 425601 (2016).

\bibitem{Saurabh} Singh S.\and Pandey S. K., Philosophical Megazine {\bf 97}, 4511 (2017).

\bibitem{Bell} Bell L. E., Science {\bf 321}, 1457 (2008).

\bibitem{Yee} Haule K., Yee C.-H. \and Kim K., Phys. Rev. B {\bf 81}, 195107 (2010).


 


\end{thebibliography}
\end{document}